# Modeling Diffusion-Limited Crystal Growth from Vapor using a Commercial Finite-Element Analysis Code

K. G. Libbrecht[1]
Department of Physics, California Institute of Technology
Pasadena, California 91125

**ABSTRACT**
This application note describes how to model diffusion-limited crystal growth from vapor using QuickField - a commercial finite-element analysis code. The crystal growth problem is cast into the format of a steady-state heat diffusion problem, which is one type of problem QuickField is designed to solve, and we derive the relevant conversion factors and correspondences between variables. We also describe three types of boundary conditions in the crystal growth problem and their corresponding boundary conditions in the heat diffusion problem. As an example, we examine the crystal growth of ice from water vapor in a background gas using a variety of mixed boundary conditions.

[This paper is also available (with better formatting and perhaps corrections) at http://www.its.caltech.edu/~atomic/publist/kglpub.htm.]

## 1. Introduction

Our interest in this problem stems from a desire to better understand the physics of snow crystal growth from water vapor, which is an interesting case study of crystal growth more generally. The growth dynamics in this case are governed by a number of factors, with vapor diffusion and attachment kinetics at the ice surface being the dominant players. While vapor diffusion is well known and calculable in principle, our understanding of the attachment kinetics controlling ice crystal growth remains quite incomplete. As a result, many observations of the morphology of ice crystals grown under different conditions remain unexplained, even at a basic, qualitative level [1].

Accurate measurements of the growth rates of ice crystals under different conditions are useful for constraining models of the growth process, and thus for investigating attachment kinetics. Most growth experiments, however, are done in the presence of at least some background gas, in which case the growth rates are also affected by diffusion. To address questions about attachment kinetics using growth measurements, it is thus necessary to model the effects of vapor diffusion. Commercial finite-element codes are useful for this purpose, since only the simplest diffusion problems can be solved analytically.

We have been using one such code in particular – QuickField, by Tera Analysis [2] – and find it quite useful for quick and easy diffusion modeling problems. The code is written for heat diffusion, however, so a number of conversion factors are necessary for quantitative modeling of ice crystal growth. The purpose of this document is to provide these conversion factors along with a quick guide to using the code for modeling crystal growth. While the factors are specific to this particular

---

[1] e-mail address: kgl@caltech.edu

code, the general methodology could be applied to other commercial codes as well.

## 2. Notation and Conversions

Following the notation of [1] and [3], we write the growth velocity normal to the surface in terms of the Hertz-Knudsen formula

$$v_n = \alpha \frac{c_{sat}}{c_{solid}} \sqrt{\frac{kT}{2\pi m}} \sigma_{surf} \qquad (1)$$
$$= \alpha v_{kin} \sigma_{surf}$$

where the latter defines the velocity $v_{kin}$. In this expression $kT$ is Boltzmann's constant times temperature, $m$ is the mass of a water molecule, $c_{solid} = \rho_{ice}/m$ is the number density for ice, $\sigma_{surf} = (c_{surf} - c_{sat})/c_{sat}$ is the supersaturation just above the growing surface, $c_{surf}$ is the water vapor number density at the surface, and $c_{sat}(T)$ is the equilibrium number density above a flat ice surface. The parameter $\alpha$ is known as the *condensation coefficient*, and it embodies the surface physics that governs how water molecules are incorporated into the ice lattice, collectively known as the *attachment kinetics*. The attachment kinetics can be nontrivial, so in general $\alpha$ will depend on $T$, $\sigma_{surf}$, as well as the surface structure and geometry, surface chemistry, etc. If molecules that strike the surface are instantly incorporated into it, then $\alpha = 1$; otherwise we must have $\alpha \leq 1$. The appearance of crystal facets indicates that the growth is limited by attachment kinetics, so we must have $\alpha < 1$ on faceted surfaces. For a molecularly rough surface, or for a liquid surface, we expect $\alpha \approx 1$. Experiments with ice growing from vapor are nearly always in a near-equilibrium regime, where $\sigma_{surf} \ll 1$.

Ice growth in air is actually a double diffusion problem – particle diffusion describing the influx of molecules through the air, and heat diffusion describing the outflow of latent heat generated by solidification. In many cases we can ignore heat diffusion, however. If the growth is on a substrate, then the latent heat is efficiently carried away via conduction through the ice and substrate [1]. If no substrate is present, then typically the crystal temperature rises slightly from the release of latent heat, and this can be modeled by a simple change in $c_{sat}$ [1]. Therefore, from this point we will consider only the particle diffusion problem.

Particle transport is described by the diffusion equation

$$\frac{\partial c}{\partial t} = D\nabla^2 c \qquad (2)$$

where $c(x)$ is the water molecule number density surrounding the crystal and $D$ is the diffusion constant. The timescale for diffusion to adjust the vapor concentration in the vicinity of a crystal is $\tau_{diffusion} \approx R^2/D$, where $R$ is a characteristic crystal size. This should be compared with the growth time, $\tau_{growth} \approx 2R/v_n$, where $v_n$ is the growth velocity of the solidification front normal to the surface. The ratio of these two timescales is called the Peclet number, $p = Rv_n/2D$. For typical growth rates of snow crystals we have $p \sim 10^{-5}$, which means that diffusion adjusts the particle density around the crystal much faster than the crystal shape changes. In this case the diffusion equation reduces to Laplace's equation

$$\nabla^2 c = 0 \tag{3}$$

which must be solved with the appropriate boundary conditions. Using this slow-growth limit simplifies the problem considerably in comparison to much of the literature on diffusion-limited solidification.

The continuity equation at the interface gives

$$v_n = \frac{D}{c_{solid}}\left(\hat{n}\cdot\vec{\nabla}c\right)_{surf} = \frac{c_{sat}D}{c_{solid}}\left(\hat{n}\cdot\vec{\nabla}\sigma\right)_{surf} \tag{4}$$

where $\sigma(x) = [c(x) - c_{sat}]/c_{sat}$ is the supersaturation field and $c_{sat}$ is independent of spatial position (because we are considering the isothermal case).

In solving Laplace's equation, we typically specify the boundary condition far from the crystal as $\sigma = \sigma_\infty$. The boundary conditions at the crystal surface, however, can be specified in any of three different ways. The first choice is to specify the surface supersaturation $\sigma_{surf}$ over all or part of the crystal. If $\alpha \approx 1$ and $v/v_{kin} \ll \sigma_\infty$, then Equation (1) suggests that $\sigma_{surf} \approx 0$ is a reasonable approximation for the surface condition. For most ice prisms, however, $\alpha \ll 1$ on the facets and $\sigma_{surf} \approx 0$ is not a good approximation. In such cases, one does not know $\sigma_{surf}$ on the surface, so a different choice of boundary conditions is necessary.

The second possibility is to specify $\alpha \sim \sigma^{-1}\nabla_n\sigma$ over all or part of the crystal surface. This is often impractical for crystal prisms, especially when comparing with observations, because $\alpha$ usually varies considerably over the surface, including over a single facet surface.

The third choice for boundary conditions is to specify the growth velocity $v_n \sim \nabla_n\sigma$ at the surface. This choice is especially useful for comparing with experiments on prism growth, since $v_n$ is a measured quantity. Furthermore, as long as the crystal growth morphology remains that of a simple hexagonal prism, then we must have that $v_n = $ *constant* over an entire facet. For this reason, specifying $v_n$ is often the most practical choice of boundary conditions at the ice surface.

## 3. Using QuickField

For steady-state heat diffusion we have the equation

$$k\nabla^2 T = -Q \tag{5}$$

where $k$ is the thermal conductivity (with MKS units W m$^{-1}$ K$^{-1}$), $T$ is temperature, and $Q$ describes heat sources per unit volume (with MKS units W m$^{-3}$). In the absence of heat sources this reduces to Laplace's equation for the temperature, $\nabla^2 T = 0$. In addition, we have the heat flux

$$F = -k\nabla T \tag{6}$$

where $F$ has MKS units of W m$^{-2}$.

To use QuickField for particle diffusion problems, we use the correspondences

$$T \text{ (K)} \to \sigma \text{ (absolute)} \tag{7}$$
$$F \text{ (W m}^{-2}) \to -v \text{ (}\mu\text{m/sec)} \tag{8}$$

where $\sigma$ is the supersaturation and $v$ is the growth velocity. By absolute units for $\sigma$, we mean that a supersaturation of one percent means $\sigma = 0.01$. Note that the heat flux perpendicular to a surface, not necessarily the total heat flux, corresponds to the perpendicular growth velocity of the surface.

For the thermal conductivity $k$ we have the correspondence

$$k \to 10^6 \frac{c_{sat} D}{c_{solid}} \tag{9}$$

For our example problem of ice growth in air at -15 C and a pressure of one atmosphere, we take $c_{sat}/c_{solid} = 1.5 \times 10^{-6}$, $D = 2 \times 10^{-5}$ m s$^{-2}$, and $v_{kin} = 200$ $\mu$m/sec, which gives $k = 3.0 \times 10^{-5}$ in this case. The added factor of $10^6$ comes from Equation 6 and our choice to measure $v$ in $\mu$m/sec. With these conventions we then apply boundary conditions as follows:

**Constant Supersaturation at a Boundary**. Here we simply set $T$ at a surface equal to $\sigma_{surf}$. For example, the far-away boundary is usually specified as $T = \sigma_\infty$ in absolute units.

**Growth Velocity at a Boundary**. Here we use the heat flux boundary condition in QuickField with $q = -v$, where $v$ is specified in $\mu$m/sec. This boundary condition is most useful when the growth velocity of a facet is a measured quantity, since then $v$ is constant across the facet surface, while $\alpha$ and $\sigma$ generally vary across a facet surface.

**Condensation Coefficient at a Boundary**. In this case we use the convective boundary condition in QuickField. The relation $F = \alpha_Q \Delta T$, where $\alpha_Q$ is the QuickField parameter, corresponds to the crystal growth equation $v = \alpha v_{kin} \sigma$, so we use the correspondence

$$\alpha_Q \to \alpha v_{kin}$$

with $v_{kin}$ in $\mu$m/sec. For growth at -15 C, this becomes $\alpha_Q = 200\alpha$.

When starting a new QuickField problem, we choose a steady-state heat transfer problem and pick our units to be microns. We then draw the crystal, specify $k$ in the space around the crystal, and specify boundary conditions as appropriate. Note that when a boundary condition is not specified, that means the surface in question is not a boundary (but may be a symmetry surface, etc.)

## 4. Testing with the Spherical Solution

The diffusion equation can easily be solved analytically for the growth of spherical crystals [3], so this allows a convenient test of the software and conversions. To this effect, we considered the growth of a spherical crystal with $R = 10$ $\mu$m inside a spherical outer boundary with $r_{out} = 300$ $\mu$m. A screen shot of the geometry is shown in Figure 1. Here the grid radius (a QuickField parameter that determines the density of the finite-element grid) was set to equal 1 $\mu$m at the inner corners and 20 $\mu$m at the outer corners. We assumed growth in air at a pressure of one atmosphere, so chose a QuickField thermal conductivity (see above) of $k = 3.0 \times 10^{-5}$.

## 4.1. Fast Kinetics

If the surface kinetics are fast in comparison to diffusion (the $\alpha_{diff} \to 0$ limit, where $\alpha_{diff}$ is defined below and in [3]), then the inner boundary has $\sigma_R = 0$ and we take the outer boundary to have $\sigma = \sigma_{out}$. The diffusion equation then has the solution

$$v = \frac{c_{sat}}{c_{solid}} \frac{D}{R} \left(1 - \frac{R}{r_{out}}\right)^{-1} (\sigma_{out} - \sigma_R)$$

$$= 3.103 \, \mu\text{m/sec}$$

where the numerical value assumes $\sigma_{out} = 1$ and the values for the other parameters as given above. Note the $\left(1 + \frac{R}{r_{out}}\right)$ factor that comes from the fact that our outer boundary is not at infinity.

The results from QuickField are shown in the screen shot in Figure 2. Putting the "voltmeter" at various points just above the surface of the sphere gives velocities (from the reported heat flux $F$) of $2.8 \pm 0.1$ $\mu$m/sec. The error in the numerical result is mainly from the finite grid size. Changing the inner grid radius to 0.1 $\mu$m and the outer grid radius to 2 $\mu$m produces a growth velocity of $3.075 \pm 0.01$ $\mu$m/sec, which agrees with the theoretical value to about one percent.

## 4.2. Finite Kinetics

When $\alpha$ or $v$ is defined at the surface and $\sigma_R$ is not, the spherical solution gives

$$v = \alpha v_{kin} \sigma_R$$

$$= \frac{c_{sat}}{c_{solid}} \frac{D\sigma_{out}}{R} \left[\frac{\alpha}{\alpha(1 - R/r_{out}) + \alpha_{diff}}\right]$$

where

$$\alpha_{diff} = \frac{c_{sat}}{c_{solid}} \frac{D}{v_{kin} R}$$

$$= \frac{D}{R} \sqrt{\frac{2\pi m}{kT}}$$

which gives $\alpha_{diff} = 0.015$ using the parameters given above. We used this analytic solution with $\sigma_{out} = 1$ to generate a range of specific test solutions having $(\alpha, v) = (0.1, 2.6866), (0.01, 1.216), (0.001, 0.1879)$, where the growth velocity is in $\mu$m/sec. We tested the code two ways - first by setting $\alpha$ (using convective boundary conditions) and determining $v$ numerically, and again by setting $v$ (using heat flux boundary conditions) and determining $\alpha$ numerically using the relation $v = \alpha v_{kin} \sigma_R$ (extracting $\sigma_R$ from the numerical solution). In both cases the code produced values that were accurate to about 0.3 percent when we used the finer grid described above.

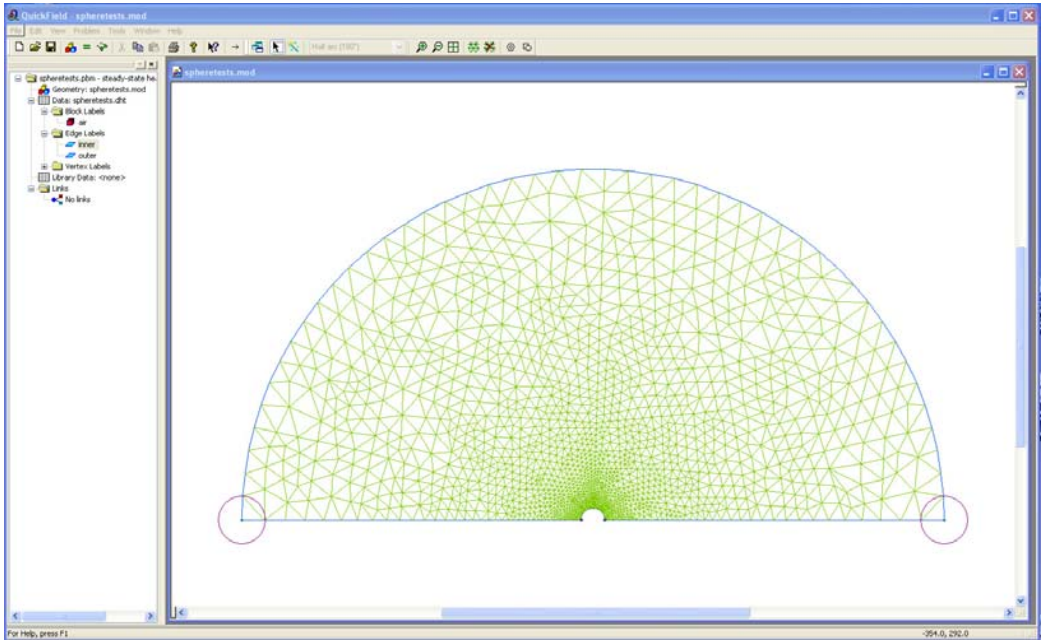

Figure 1. Display in QuickField showing the space between the growing ice crystal (inner half-circle) and the outer boundary (outer half-circle), together with the finite-element grid. Here the (z,r) coordinate axes lie along the horizontal and vertical axes, respectively.

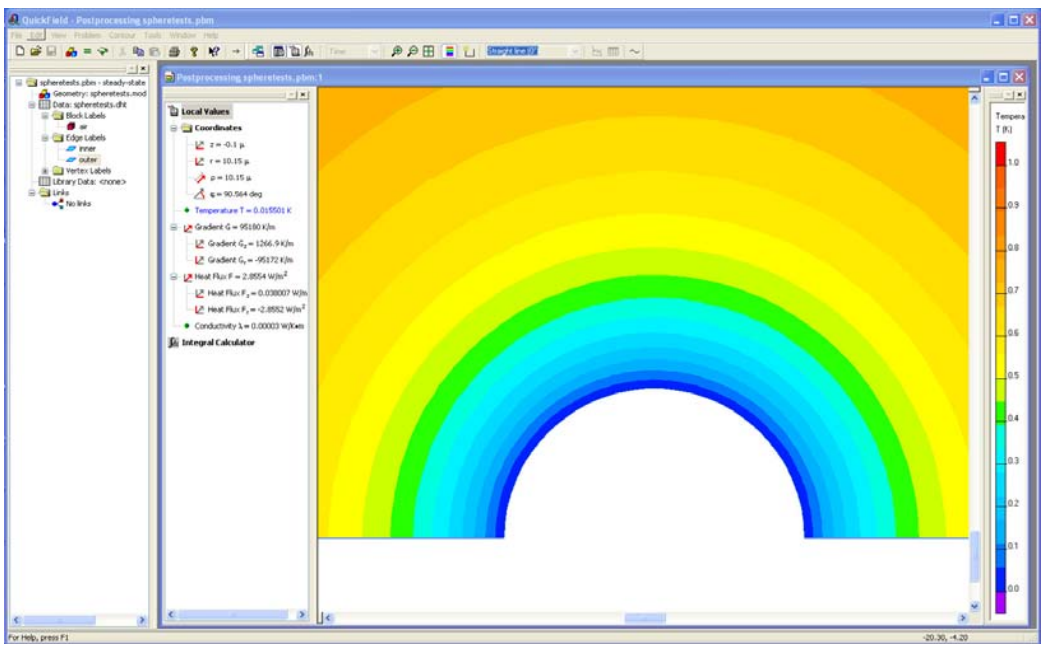

Figure 2. QuickField solution display, in "voltmeter" mode for sampling the calculated temperature (corresponding supersaturation) and heat flux (giving the growth velocity) around the crystal. The scale is blown up with respect to the previous plot.